\documentclass[conference]{IEEEtran}
\IEEEoverridecommandlockouts
\usepackage{cite}
\usepackage{amsmath,amssymb,amsfonts}
\usepackage{algpseudocode}
\usepackage{graphicx}
\usepackage{textcomp}
\usepackage{xcolor}
\usepackage{algorithm}
\usepackage[font=footnotesize,skip=2pt]{caption}
\def\BibTeX{{\rm B\kern-.05em{\sc i\kern-.025em b}\kern-.08em
    T\kern-.1667em\lower.7ex\hbox{E}\kern-.125emX}}

\title{Learning-to-Optimize Graph Framework for Real-Time Relay-Link Optimization in NR-V2X\\
{\footnotesize %\textsuperscript{*}Note: Sub-titles are not captured for https://ieeexplore.ieee.org  and}
\thanks{This work was supported by the European Union under the Italian National
Recovery and Resilience Plan (NRRP) of NextGenerationEU, partnership on
“Telecommunications of the Future” (PE00000001 - program “RESTART”).}
}

\author{\IEEEauthorblockN{
Giambattista Amati, Federica Mangiatordi, Simone Angelini, Emiliano Pallotti, Pierpaolo Salvo}

\textit{Fondazione Ugo Bordoni}\\
Rome, Italy \\
Emails: \{gamati, fmangiatordi, sangelini, epallotti, psalvo\}@fub.it}
}
\begin{document}
\maketitle

\begin{abstract}
Reliable and low-latency uplink connectivity in NR-V2X requires fast relay-link optimisation that scales with highly dynamic vehicular topologies. Although Mixed-Integer Linear Programming (MILP) yields optimal multi-hop configurations, its runtime scales rapidly with graph density, limiting real-time applicability.
This paper presents a Learning-to-Optimise (L2O) framework based on Graph Isomorphism Networks (GINs) that approximates MILP relay decisions with near-constant inference latency. A dataset of 77{,}500 SUMO-GEMV\(^2\)/Sionna V2X snapshots is labelled using a comprehensive MILP oracle that enforces flow, capacity, and cycle-prevention constraints. The GIN encoder extracts topology and radio-aware embeddings, which an edge classifier maps to relay-activation probabilities.
Experimental results show that the proposed L2O model attains 0.8936 accuracy and 0.9110 recall at a 0.5 threshold, with recall increasing to 0.9704 at a relaxed 0.30 threshold, thereby preserving the MILP-selected multi-hop paths. Inference latency remains \(3\)–\(8\)\,ms, significantly outperforming MILP’s tens- to hundreds-of-milliseconds latency. At the system level, L2O achieves a 5.84\% connectivity improvement compared to 9.31\% for MILP, while still operating in real time.
These findings show that GIN-based L2O is a scalable, practical alternative to MILP for real-time relay selection in dense NR-V2X scenarios.
\end{abstract}


\section{Introduction}

NR-V2X connectivity is essential for autonomous mobility, cooperative
perception, and safety-critical vehicular applications. Modern Connected and
Automated Vehicles (CAVs) integrate heterogeneous onboard sensors
(e.g., LiDAR, radar, cameras) that require continuous high-rate uplink toward
the infrastructure \cite{ahangar2021survey,wang2018networking}. While DSRC and legacy C-V2X systems are
limited by constrained spectrum and support only basic V2X services, emerging
5G-Advanced and 6G NR-V2X architectures will offer substantially higher
bandwidth, enabling enhanced V2X (e-V2X) services that demand high capacity,
ultra-low latency, and near-perfect reliability \cite{3gpp-22-186}.

Operating under such stringent requirements, however, makes vehicular links
highly vulnerable to blockage and intermittent propagation
\cite{Sun2018PropagationMA,9119160}, thereby increasing the importance of
relay mechanisms to sustain stable connectivity. Relaying—via neighbouring
CAVs or RSUs—can effectively mitigate blockage and extend infrastructure reach
\cite{10304089}, but requires frequently updated network information and
low-latency decision-making, making classical optimisation approaches difficult
to apply in dynamic vehicular scenarios.

Relay-link optimisation must simultaneously account for radio constraints, QoS requirements, RSU bandwidth, and multi-hop flow conservation, while ensuring that every CAV maintains a feasible uplink path to the core network. The underlying problem is NP-hard, and even when a
Mixed-Integer Linear Programming (MILP) formulation admits an optimal solution, its computation time scales superlinearly with network size, making MILP
infeasible for real-time operation in dense urban settings.

A promising alternative is \emph{Learning-to-Optimise} (L2O), where neural
models learn structural regularities from optimal solutions and replace slow
optimisation with fast inference \cite{Chen2024,shao2021learningrobustcombinatorialoptimization}. Graph Neural Networks (GNNs) are
particularly well suited to this task because NR-V2X systems are naturally
graph-structured, with nodes representing vehicles and RSUs and edges encoding
time-varying link opportunities \cite{10274765}. Message-passing architectures capture spatial,
topological, and radio-aware dependencies that conventional feature-flattening
approaches fail to exploit.

In this study, we introduce a GIN-based L2O framework for multi-hop relay-link optimisation in NR-V2X. Exploiting the representational strength of Graph Isomorphism Networks, the model learns topology-aware embeddings that approximate MILP-level relay decisions while supporting real-time inference in large-scale vehicular networks.

\section{Related Work}

Early attempts to optimise vehicular communications using neural networks relied on fully-connected architectures trained as function approximators to emulate classical power allocation or interference management algorithms \cite{8891446, 8943940}. In such methods, the input space was typically reduced to a flat feature vector containing channel gains or signal-to-noise ratios, while topological information was implicitly ignored. As a result, these approaches usually scale poorly as the communication graph becomes dense and dynamic, and they exhibit limited generalisation when channel conditions or vehicular density deviate from the training configuration.

A first step toward more expressive representation learning has been to apply Graph Neural Networks (GNNs) to vehicle-to-vehicle (V2V) and vehicle-to-infrastructure (V2I) resource management \cite{9322537,ji2025graphneuralnetworksdeep,amati2025AEIT,HUAN20255427}. These works have shown that message-passing mechanisms can capture structural dependencies between vehicles and base stations, thereby enabling learning-based resource allocation in settings characterised by interference coupling or dual-mode connectivity. In this context, GNNs have been employed to approximate power control, spectrum sharing, and link selection policies, thereby improving scalability relative to conventional deep neural networks.

More recently, studies have explicitly modelled edge-level relationships by introducing link embedding strategies, such as constructing line graphs to convert edge decisions into node-classification tasks \cite{9979700,naderializadeh2021wirelesslinkschedulinggraph,10068338,jiang2020coembeddingnodesedgesgraph, Amati2024AEIT}. While these techniques provide a richer representation of link characteristics---including capacities, interference, and directionality---they often suffer from increased computational complexity and memory usage, which may limit applicability in dense vehicular deployments.

Additional progress has been reported in graph-based handover prediction, topology-aware mobility management, and distributed scheduling, often combining GNN-based encoders with reinforcement learning or heuristic decision-making modules \cite{11096201,tang2025varisacv2xassuredconnectivity,11268640,bermudez2025graph}. These approaches highlight the potential of graph learning to exploit spatio-temporal patterns in dynamic mobile networks. However, they typically focus on partial tasks (e.g., V2V link adaptation, handover triggering, or distributed interference management) rather than end-to-end relay optimisation under strict NR-V2X constraints.

In contrast to the above, our work contributes a unified formulation explicitly tailored to multi-hop NR-V2X relay selection and connectivity maximisation. In particular, we: (i) define a comprehensive Mixed-Integer Linear Programming (MILP) model for optimal multi-hop relay decisions under realistic flow, capacity, and cycle-prevention constraints; (ii) train a supervised Graph Isomorphism Network (GIN) based Learning-to-Optimize (L2O) surrogate that learns to approximate MILP solutions at inference time; and (iii) validate the resulting L2O framework over 15{,}500 simulated graph instances generated from realistic SUMO--GEMV$^2$/Sionna vehicular environments. To the best of our knowledge, this is among the first works to combine MILP-based optimality with GIN-driven Learning-to-Optimise at scale for relay selection in NR-V2X networks.
%%%%%%%%

\section{Graph Representation}
Each V2X snapshot corresponds to the instantaneous communication topology of the
vehicular system and is modelled as a directed graph $G_t=(V_t,E_t)$. The node
set includes both CAVs and RSUs, i.e., $V_t = V_{\text{CAV}} \cup V_{\text{RSU}}$,
while directed edges $(i,j)\in E_t$ represent feasible wireless links satisfying
the NR-V2X SNR threshold.

Nodes are characterised by spatial coordinates, a CAV/RSU role indicator, and—%
for CAVs—the current uplink traffic demand associated with 3GPP NR-V2X services
(e.g., infotainment or cooperative perception). Additional mobility attributes
such as speed and heading are also included. Edges carry radio-aware features,
including SNR, SINR, LoS/NLoS classification, and the Shannon capacity computed
from the instantaneous SINR.
As vehicles move and channel conditions change, the communication graph
changes accordingly. We model the complete network trace as a time-indexed
sequence $\{G_t\}_{t=0}^{T-1}$ sampled every 1~s, forming a dynamic, physically grounded basis for learning multi-hop connectivity and relay selection strategies.

\subsection{Simulation Framework}
Vehicular mobility was generated using SUMO \cite{SUMO2,SUMO} on top of an
OpenStreetMap (OSM) topology \cite{OpenStreetMap} representing a realistic
$1\,\text{km}^{2}$ urban area of Rome (Porta Pia). The OSM-derived geometry
preserves actual road layout, intersection density, and lane structure, enabling
heterogeneous and representative mobility patterns.

The SUMO-generated CAV positions were then fed into a hybrid GEMV$^2$/Sionna
radio simulator \cite{GEMV2,sionna}, which models geometry-aware path loss,
LoS/NLoS conditions, shadowing, multipath, and simplified ray tracing, ensuring
PHY-layer consistency with 3GPP NR-V2X assumptions. In this integrated framework,
SUMO provides mobility and link distances, while Sionna computes time-varying
propagation attributes.
Radio simulations adopt the following parameters: RSU transmit power 10~dBm,
vehicle transmit power 0~dBm, antenna gains 1~dBi, dipole antenna pattern,
carrier frequency 5.9~GHz, and 800~MHz bandwidth.
At each timestamp, the combined mobility–radio pipeline produces a complete V2X
snapshot including node positions, feasible links, and propagation metrics,
ensuring that edge feasibility in $G_t$ arises from physically grounded urban
mobility and channel conditions rather than synthetic assumptions.

\subsection{Dataset Generation}

A single large-scale dataset was generated from the $1\,\text{km}^{2}$ Porta Pia
scenario to train and validate the proposed Learning-to-Optimise (L2O)
surrogate. The dataset consists of 77{,}500 labelled V2X graph instances
$\{G_t\}$ sampled at $1$\,s intervals, each capturing a distinct vehicular
configuration together with the corresponding radio conditions derived from the
GEMV$^2$/Sionna pipeline.

Each snapshot is processed through the Mixed-Integer Linear Programming (MILP)
formulation described in Sec.~\ref{lab_optimization_problem}, used as an
\textit{oracle} that computes the optimal multi-hop relay configuration. The
resulting optimal activation matrix is stored as edge-level supervision and
constitutes the ground-truth target for the L2O training. In this way, the
dataset provides exact optimal solutions that would otherwise be computationally
expensive to obtain at inference time.

To assess generalisation capabilities, the vehicular density is systematically
varied across the simulation horizon, producing graphs with heterogeneous node
populations and connectivity levels. The generated topologies range from sparse
low-traffic conditions to highly congested urban settings, resulting in a
diverse set of graph structures with varying numbers of CAVs, RSUs, and
candidate radio links. This diversity is fundamental to preventing overfitting
towards a specific mobility pattern or connectivity regime.

Overall, the dataset forms a fully labelled benchmark for NR-V2X relay
optimisation in dense urban environments, enabling supervised training of
graph-based surrogate models while preserving physical realism and MILP-grade
optimality.
The dataset is randomly partitioned into training, validation, and test sets
according to a standard 70\%--10\%--20\% split, ensuring fair model evaluation
and preventing temporal or spatial leakage across subsets.

\section{MILP Oracle for Relay-Link Optimisation}
\label{lab_optimization_problem}

The optimal relay setup is determined by solving a Mixed-Integer Linear Programming (MILP) problem that models radio, flow, and capacity constraints in NR-V2X multi-hop links. The MILP formulation serves as an ``oracle''
that produces optimal link activations for each V2X snapshot. 
\subsection{Decision Variables}

To capture connectivity decisions and traffic routing, the following decision
variables are defined:
\begin{itemize}
    \item $v_{ij}\in\{0,1\}$: binary variable indicating whether the directed
    link $(i,j)$ is activated for relaying.
    \item $f_{ij}\ge0$: continuous variable denoting the amount of traffic flow
    routed over link $(i,j)$.
    \item $z_i\in\{0,1\}$: binary variable indicating whether CAV~$i$ is
    successfully connected to the infrastructure (either directly or via
    multi-hop relays).
    \item $u_i\in\{0,1,\dots,M-1\}$: integer ordering variable used to eliminate
    cycles in the CAV relay subgraph, where $M$ is an upper bound on the number
    of CAVs.
    \item $w_{ik}\in\{0,1\}$: auxiliary binary variable used to linearise
    products of the form $v_{ik}\cdot z_k$ and enforce relay consistency.
\end{itemize}
\subsection{Objective}
The goal of the optimisation is to maximise the number of CAVs that maintain a
valid uplink path towards at least one RSU, possibly through intermediate CAV
relays. This is formalised by the objective function
\begin{equation}
    \max \sum_{i\in\text{CAV}} z_i.
\end{equation}

The objective is to derive viable paths from the maximum possible number of CAVs to any accessible RSU, subject to prevailing radio and data-flow constraints.

\subsection{Key Constraints}

\noindent\textbf{Flow conservation:}
\begin{equation}
\sum_j f_{ij} - \sum_k f_{ki} = d_i z_i, \quad \forall i \in \text{CAV},
\end{equation}
where $d_i$ denotes the uplink data demand of CAV~$i$. Activated nodes must
provide sufficient outgoing flow to reach an RSU.

\medskip
\noindent\textbf{RSU capacity:}
\begin{equation}
\sum_i f_{ij} \le M_{RSU_j}, \quad \forall j \in \text{RSU},
\end{equation}

where $M_{\text{RSU}_{j}}$ denotes the maximum aggregate data that an RSU$_j$ can serve,
which depends on the amount of available radio resources (e.g., resource blocks)
assigned to V2X traffic.

\medskip
\noindent\textbf{Link capacity:}
\begin{equation}
f_{ij} \le C_{ij} v_{ij},
\end{equation}
where $C_{ij}$ is the maximum capacity of the wireless link $(i,j)$,
derived from the associated SINR estimate.

\medskip
\noindent\textbf{CAV connectivity:}
Each CAV can either be disconnected ($z_i=0$) or connected with exactly one
outgoing relay link ($z_i=1$), identifying its next hop (either another CAV or
an RSU):
\begin{equation}
    \sum_{j : (i,j) \in \mathcal{E}_{\text{valid}}} v_{ij} = z_i,
    \qquad \forall i \in \text{CAV}.
\end{equation}
This constraint enforces single-path connectivity and prevents multiple outgoing
relay selections for a given CAV.

\medskip
\noindent\textbf{Relay consistency:}
To ensure that a CAV $i$ may rely on another CAV $k$ only if $k$ is itself
connected, auxiliary variables $w_{ik}$ are used to linearise the product
$v_{ik} z_k$. For each valid relay edge $(i,k)\in\mathcal{E}_{\text{valid}}$,
\begin{align}
    w_{ik} &\le v_{ik}, &
    w_{ik} &\le z_k, \\
    w_{ik} &\ge v_{ik} + z_k - 1, &
    z_i    &\ge w_{ik}.
\end{align}
These constraints impose $w_{ik}=v_{ik}z_k$ and propagate connectivity along
multi-hop paths, ensuring that a relay can serve upstream CAVs only if it is
itself connected.

\medskip
\noindent\textbf{Cycle elimination:}
To prevent cycles among relaying CAVs, we assign an ordering variable $u_i$ to each CAV $i$ and apply Miller--Tucker--Zemlin (MTZ) constraints on CAV--CAV edges. For each feasible CAV--CAV link $(i,j)$, we impose
\begin{equation}
    u_i + 1 \leq u_j + M \left(1 - v_{ij}\right),
    \qquad \forall (i,j) \in \mathcal{E}_{\text{CAV--CAV}},
\end{equation}
where $M$ is an upper bound on the number of CAVs and $\mathcal{E}_{\text{CAV--CAV}}$
is the set of candidate CAV--CAV relay links. With appropriate bounds on $u_i$, these constraints ensure that the selected CAV--CAV links form a directed acyclic structure, thus yielding valid multi-hop routes.

\subsection{Optimal Activation Matrix (Oracle)}
The MILP solution yields, for each V2X snapshot, the optimal binary link activation matrix $\mathbf{A}^\star=\{A_{ij}^\star\}$. The following condition defines the optimality of the latter: if $A_{ij}^\star$ is set to $1$ then link $i$ and $j$ is selected in the optimal multi-hop configuration.
The matrix under consideration serves as the oracle output and is used as ground-truth supervision for training the proposed GIN-based L2O model. During the learning process, the surrogate is exposed to optimal relay decisions, thereby progressively internalising the underlying structure. At inference time, this enables real-time prediction of relay activations without requiring repeated invocations of the MILP solver. This approach circumvents the computational limitations of the MILP solver while maintaining the integrity of the decision-making process.

\section{GIN-Based Topology Encoding}
\label{sec:gin_encoder}

The relay-activation framework employs a Graph Isomorphism Network encoder to derive topology-aware node embeddings from each V2X graph snapshot. It maps raw node features to latent representations that capture the structural context of vehicles and RSUs, incorporating multi-hop neighbourhood information. These embeddings serve as the foundation for edge-level activation decisions in the Learning-to-Optimise (L2O) architecture (Sec.~\ref {sec:learning2optimize}).
\begin{figure}[ht]
\centerline{\includegraphics[width=0.999\linewidth]{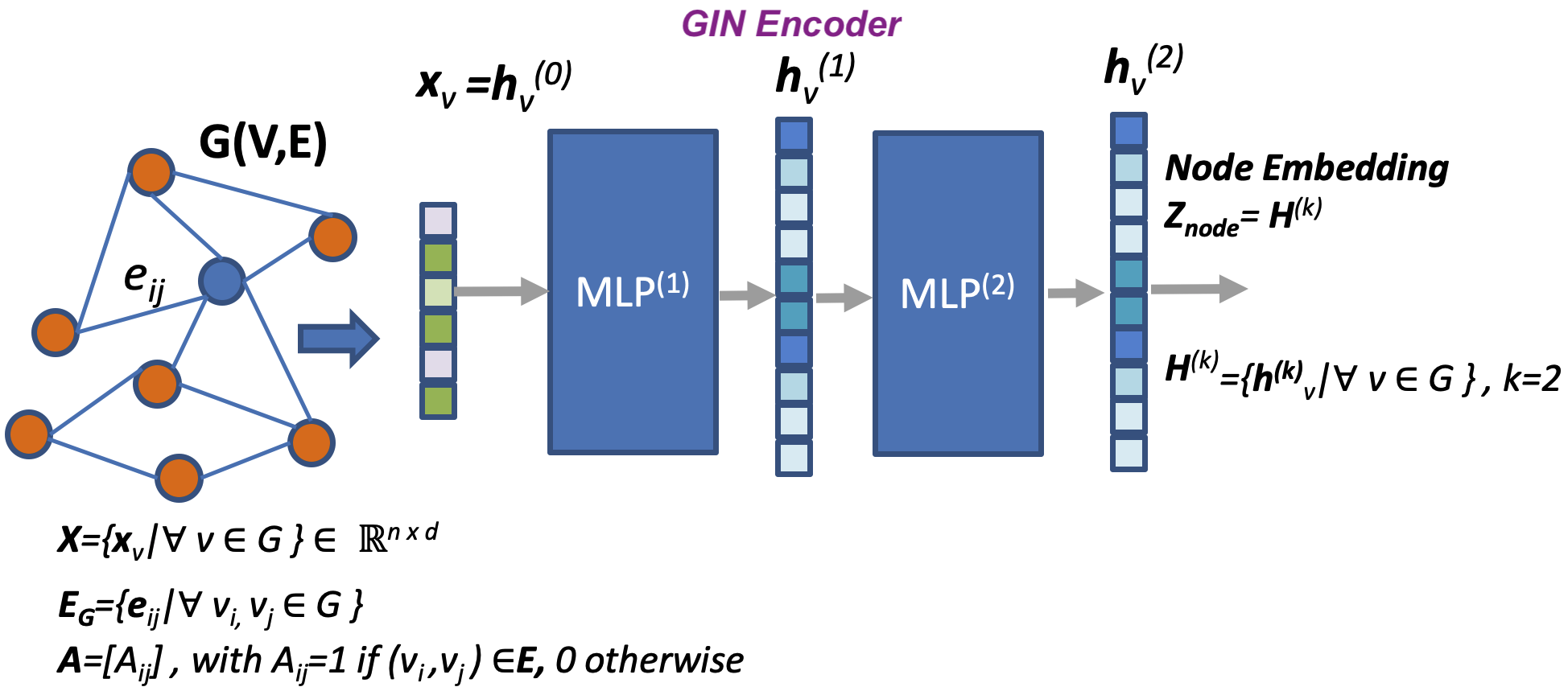}}
\caption{GIN for V2X topology encoding}
\label{figL2O}
\end{figure}

GIN is adopted due to its strong discriminative capability: it is theoretically
as expressive as the Weisfeiler–Lehman (WL) graph isomorphism test
\cite{xu2019powerfulgraphneuralnetworks}, enabling it to distinguish between
nodes that share similar attributes but play different structural roles (e.g.,
isolated CAVs, relay candidates, or RSUs acting as bottlenecks). This property
is particularly important in NR-V2X, where vehicular topologies are highly
dynamic and heterogeneous. Compared with classical convolutional GNNs such as
GCN, GIN provides superior representational power for capturing subtle
relay-relevant patterns in dense multi-hop connectivity graphs.

\subsection{Message Passing}
At each propagation layer $k$, node embeddings are updated according to
\begin{equation}
h_v^{(k)} = \mathrm{MLP}^{(k)} \!\left(
(1+\epsilon^{(k)})\,h_v^{(k-1)}
+ \sum_{u\in\mathcal{N}(v)} h_u^{(k-1)}
\right),
\end{equation}
where $h_v^{(k)}$ denotes the embedding of node $v$ at layer $k$,
$\mathcal{N}(v)$ represents the one-hop neighbourhood of $v$, and
$\epsilon^{(k)}$ is a learnable scalar. The summation operator aggregates
information from neighbouring nodes, while the residual term
$(1+\epsilon^{(k)})\,h_v^{(k-1)}$ preserves node identity and improves gradient
flow.
The use of injective multiset aggregation via \textit{SUM} ensures that distinct
neighbourhood structures map to distinct embedding distributions, enabling the
encoder to distinguish topologically different communication configurations
even in dense and partially symmetric urban scenarios.

%For downstream tasks requiring global information, a graph-level representation can be obtained through simple sum pooling:
%\begin{equation}
%Z_{\text{graph}} = \sum_{v\in V} h_v^{(K)},
%\end{equation}
%where $K$ denotes the final GIN layer. Although the present study focuses on
%edge-wise link activation rather than graph classification, the global
%embedding may be exploited for future extensions involving temporal forecasting or RL-based topology control.
The resulting node embeddings $\{h_v^{(K)}\}$ represent the potential connectivity of each node. An edge classifier can score each candidate link by concatenating the embeddings of its endpoints with the corresponding edge features. This enables relay activation decisions to be informed by local radio conditions and the V2X topology.

\section{GIN-Based Architecture for Relay-Link Activation}
\label{sec:learning2optimize}
The proposed L2O surrogate consists of two tightly coupled components:  
(i) a Graph Isomorphism Network (GIN) encoder that computes topology-aware node
embeddings, and (ii) an edge-level classifier that predicts relay activation
probabilities.

\subsection{Node and Edge Features}

Each V2X snapshot $G_t = (V_t,E_t)$ is enriched with node and edge attributes
reflecting geometric, functional, and radio-layer properties.  
Node features encode spatial position, node type, and traffic demand:
\begin{equation}
x_v = [x_v,\, y_v,\, \text{CAV flag},\, d_v],
\end{equation}
where $(x_v,y_v)$ are the planar coordinates and $d_v$ denotes the data-rate
requirement for CAVs.  
Edge features encode the instantaneous quality of each wireless link.  
For each directed edge $(u,v)$, we define the feature as the Shannon capacity based on the link SINR:
\begin{equation}
e_{uv} = C_{\text{Shannon}}(u,v)
       = B \log_2\!\left(1 + \mathrm{SINR}_{uv}\right),
\end{equation}
with $B$ denoting the available bandwidth allocated to the wireless link. This yields a physically meaningful link-quality metric that reflects the actual radio conditions between the nodes.
All features are normalised to ensure numerical stability across heterogeneous
graph instances.

\subsection{GIN Encoder}

The GIN encoder applies two message-passing layers, each using \textsc{SUM}
aggregation and an injective MLP as the update rule.  
Each MLP consists of two linear layers with intermediate {Graph Normalisation},
{ReLU activation}, and a {dropout} rate of $p=0.2$, a design that enhances
expressive capacity while improving stability and generalisation in
graph-structured data. In this context, Graph Normalisation is preferred over Batch Normalisation because it ensures consistent normalisation across graphs of varying sizes and mitigates dependence on batch statistics, which often destabilises GNNs in dynamic V2X settings.
The learnable parameter $\epsilon$ in the GIN formulation modulates the relative
weight between a node's self-embedding and information aggregated from its
neighbours, increasing representational flexibility in dense and sparse
connectivity regimes.
The encoder outputs a $256$-dimensional embedding $h_v$ for each node $v$,
capturing both local radio characteristics and higher-order connectivity patterns
that emerge from multi-hop paths.

\subsection{Edge-Level Relay Activation Classifier}

For each directed edge $(u,v)\in E_t$, the model evaluates relay suitability by
concatenating the endpoint embeddings with the corresponding edge features:
\begin{equation}
\hat{y}_{uv}
= \sigma\!\left(
    \mathrm{MLP}\big(
        [h_u \Vert h_v \Vert e_{uv}]
    \big)
\right),
\end{equation}
where $\sigma(\cdot)$ is the sigmoid function and $\Vert$ denotes vector concatenation.  
The resulting score $\hat{y}_{uv}\in[0,1]$ represents the probability that link
$(u,v)$ should be activated to support multi-hop relaying and maximise end-to-end
CAV connectivity.
\begin{figure*}[htbp]
\centerline{\includegraphics[width=0.80\linewidth]{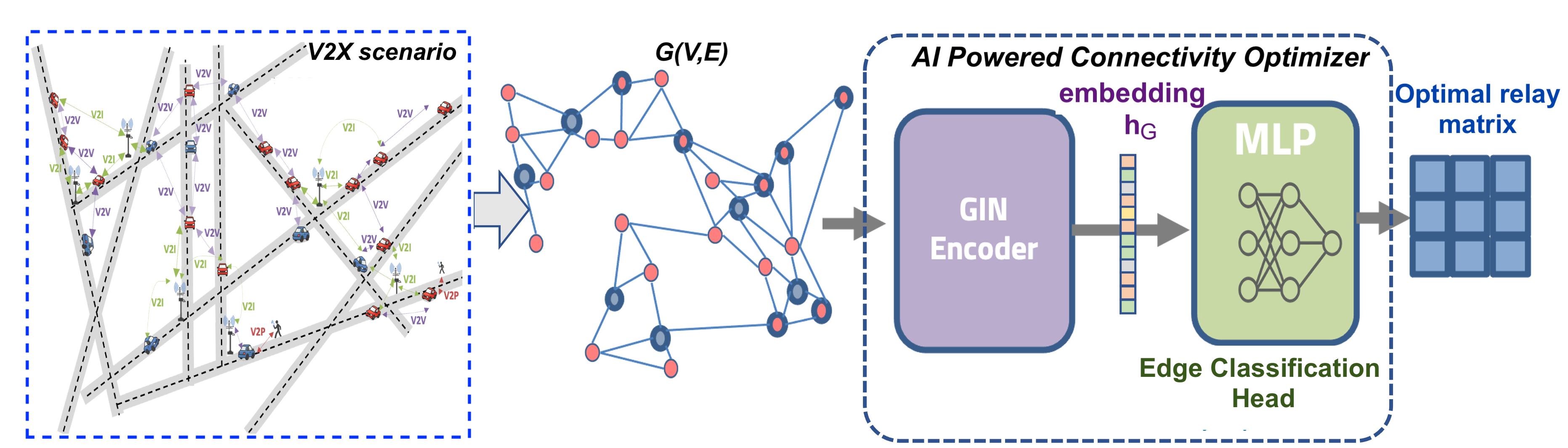}}
\caption{GIN-based L2O framework for relay activation}
\label{figL2O}
\end{figure*}
This architecture allows the model to jointly leverage geometric structure, radio conditions, and topological information, producing a powerful surrogate that can replicate MILP-level relay selection while operating at real-time inference speed.

\begin{algorithm}[th] \caption{L2O Inference %for Relay Activation 
(per snapshot $G_t$)} 
\begin{algorithmic}[1] \Require Graph $G_t=(V_t,E_t)$ %with edge features $e_{uv}$,
,threshold $\tau$ \Ensure Activated relay set $\mathcal{E}_{\text{act}}$ \State $h \leftarrow \text{GIN}(G_t;\theta_G)$ \Comment{node embeddings} \State $s_{uv}\leftarrow \sigma(\text{MLP}([h_u\Vert h_v\Vert e_{uv}];\theta_E)),\ \forall (u,v)\in E_t$ \State $\mathcal{E}_{\text{cand}}\leftarrow \{(u,v)\in E_t: s_{uv}>\tau\}$ \ForAll{$u\in V_{\text{CAV}}$} \Comment{enforce one-outgoing } \State Select $v^\star=\arg\max_{(u,v)\in \mathcal{E}_{\text{cand}}} s_{uv}$ (if any) \State Add $(u,v^\star)$ to $\mathcal{E}_{\text{act}}$ \EndFor \end{algorithmic} \end{algorithm}
\subsection{Loss Function}

The MILP-labelled V2X graphs exhibit a strong imbalance between active
(positive) and inactive (negative) relay links, with positive edges typically
representing only a small fraction of the total edge set. To obtain a stable
and recall-oriented optimisation process, the training objective is structured
as a two-stage loss combining a \emph{class-weighted Binary Cross-Entropy
(BCE) warm-up} with an \emph{Asymmetric Focal Loss (AFL)} refinement phase.

\paragraph*{BCE warm-up.}
During the initial epochs, the model is trained exclusively with a weighted BCE
loss:
\begin{equation}
\mathcal{L}_{\mathrm{BCE}}=
-\big[w_{+}y_{uv}\log p_{uv}+w_{-}(1-y_{uv})\log(1-p_{uv})\big]
\end{equation}

where $y_{uv}\in\{0,1\}$ denotes the MILP activation label for edge $(u,v)$,
and
\begin{equation}
p_{uv}=\sigma(\hat{y}_{uv}), \qquad 
\sigma(x)=\frac{1}{1+e^{-x}},
\end{equation}
is the logistic sigmoid that maps logits to probabilities.  
The weight $w_{+}$ is computed from the empirical negative-to-positive class
ratio, ensuring that rare positive links contribute proportionally to the
gradient. This stage stabilises the optimisation dynamics and prevents early
collapse due to label imbalance.

\paragraph*{Asymmetric Focal Loss refinement.}
After the warm-up phase, training transitions to a hybrid objective:
\begin{equation}
\mathcal{L}
=
\mathcal{L}_{\mathrm{BCE}}
+
\lambda \,\mathcal{L}_{\mathrm{AFL}},
\end{equation}
where $\lambda=0.5$ controls the strength of the focal refinement.  
The AFL component is defined as
\begin{equation}
\mathcal{L}_{\mathrm{AFL}}
=
- \alpha\,(1-p_{t})^{\gamma_t}
\bigl[
y_{uv}\log p_{uv}
+
(1-y_{uv})\log(1-p_{uv})
\bigr],
\end{equation}
where $p_{t}$ is the predicted probability associated with the true class.  
Asymmetry is introduced by choosing
\begin{equation}
\gamma_t =
\begin{cases}
\gamma_{\text{pos}}, & y_{uv}=1,\\[1mm]
\gamma_{\text{neg}}, & y_{uv}=0,
\end{cases}
\end{equation}
with $\gamma_{\text{pos}}=2.0$ to intensify the penalty on hard positive
samples, and $\gamma_{\text{neg}}=0.0$ to avoid excessive suppression of
negative edges. This formulation prioritises recall on the positive class,
which is essential for reconstructing MILP-grade relay paths.

\paragraph*{Joint optimisation of GIN and edge classifier.}
All parameters of the GIN encoder and of the edge-level MLP classifier are
updated jointly under the combined BCE+AFL loss. The GIN layers learn
topology-aware representations that enhance separability, while the edge MLP
learns to discriminate feasible relay links from non-activatable edges.  
This two-stage loss design consistently improves robustness across sparse and
dense vehicular scenarios and yields a surrogate model capable of approximating
MILP decisions with high fidelity.
\subsection{Training Procedure}

Training is organised into two consecutive phases designed to stabilise
optimisation under severe class imbalance and progressively enhance the
model’s ability to detect MILP-activated relay links.

\paragraph*{(i) BCE warm-up.}
During the initial stage, parameter updates are driven solely by the
class-weighted Binary Cross-Entropy loss. This warm-up phase prevents gradient
instability caused by the scarcity of positive edges and enables the model to
establish an initial separation between active and inactive links. The optimiser
is Adam with an initial learning rate of $10^{-3}$, which is gradually reduced
via cosine annealing to moderate the magnitude of early updates.

\paragraph*{(ii) Asymmetric Focal Loss refinement.}
Once stable logits have been formed, training transitions to the composite
objective $\mathcal{L}_{\mathrm{BCE}} + \lambda\mathcal{L}_{\mathrm{AFL}}$.
The Asymmetric Focal term selectively accentuates hard positive examples—those
critical for reconstructing valid multi-hop paths—while avoiding excessive
penalisation of the abundant negative edges. A step-based learning rate
schedule, with decay factor $0.85$ every 30 epochs, supports progressive
refinement of the decision boundary.

\paragraph*{Training details.}
At each iteration, a batch of 256 V2X graphs is sampled and converted into node and
edge features that encode geometry, role, traffic demand, and Shannon capacity.
Batch Normalisation and dropout (0.3--0.4) within the MLP blocks promote
robustness across heterogeneous vehicular densities and radio conditions.  
Validation is performed after every epoch to monitor loss, recall, and
activation consistency, ensuring that the GIN-based L2O model  maintains sensitivity to all
MILP-selected relay edges.

All experiments are conducted on an NVIDIA GPU to improve computational efficiency, yet the final model remains sufficiently lightweight for CPU-only inference. The joint use of BCE warm-up, AFL refinement, and adaptive learning-rate scheduling yields fast, stable convergence, enabling robust approximation of MILP relay decisions over diverse dynamic V2X topologies.

\section{Experimental results}
This section evaluates the proposed GIN-based L2O framework against the optimal
MILP oracle using: (i) link-level prediction metrics, (ii) scalability, and
runtime behaviour, and (iii) system-level connectivity gains in realistic
NR-V2X scenarios.

\subsection{Link-Level Prediction Performance}

To assess the proposed GIN-based L2O framework, we use the optimal MILP solution as the baseline. For each V2X snapshot, the MILP oracle yields the ground-truth relay-activation matrix, which is then used to derive edge-level performance metrics.

With a link-activation threshold of 0.5, the L2O framework attains, on the full dataset, an accuracy of 0.8936, precision of 0.8194, recall of 0.9110, and an F1-score of 0.8628. Similar performance is obtained on the validation subset (accuracy 0.8932, precision 0.8187, recall 0.9103, F1-score 0.8621), indicating that the model closely replicates MILP-level relay decisions.
During inference, the activation threshold can be adjusted to favour recall-focused operating points, which are essential for maintaining multi-hop paths and preventing missed MILP-triggered links. On a validation/test set of 15,500 V2X graph instances, a threshold of 0.30 achieves a high recall of 0.9704, with accuracy 0.8639, precision 0.7398, and an F1-score of 0.8395. This operating point provides a more conservative pruning strategy that favours connectivity preservation in
latency-constrained NR-V2X scenarios.

\subsection{Scalability and Runtime Analysis}
For NR-V2X scenarios with URLLC constraints, the relay-activation matrix must be
computed within a strict latency budget. In this regard, MILP and L2O exhibit
markedly different behaviours.
As shown in Fig.~3, MILP runtimes increase rapidly with graph density, ranging from a few milliseconds for small snapshots to tens or even hundreds of milliseconds for denser topologies, thereby limiting its applicability to real-time optimisation. Conversely, the proposed GIN-based L2O framework
achieves near-constant inference latency—typically $3$–$8\,\mathrm{ms}$—with no
significant dependency on the number of edges. This behaviour stems from the fixed complexity of the
forward pass, which remains compatible with URLLC timing requirements.

Overall, L2O reduces computation time by up to two orders of magnitude compared
with MILP while retaining high fidelity to the optimal solution, enabling real-time relay selection in dense and dynamic vehicular environments.
\begin{figure}[htbp]
\centerline{\includegraphics[width=0.90\linewidth]{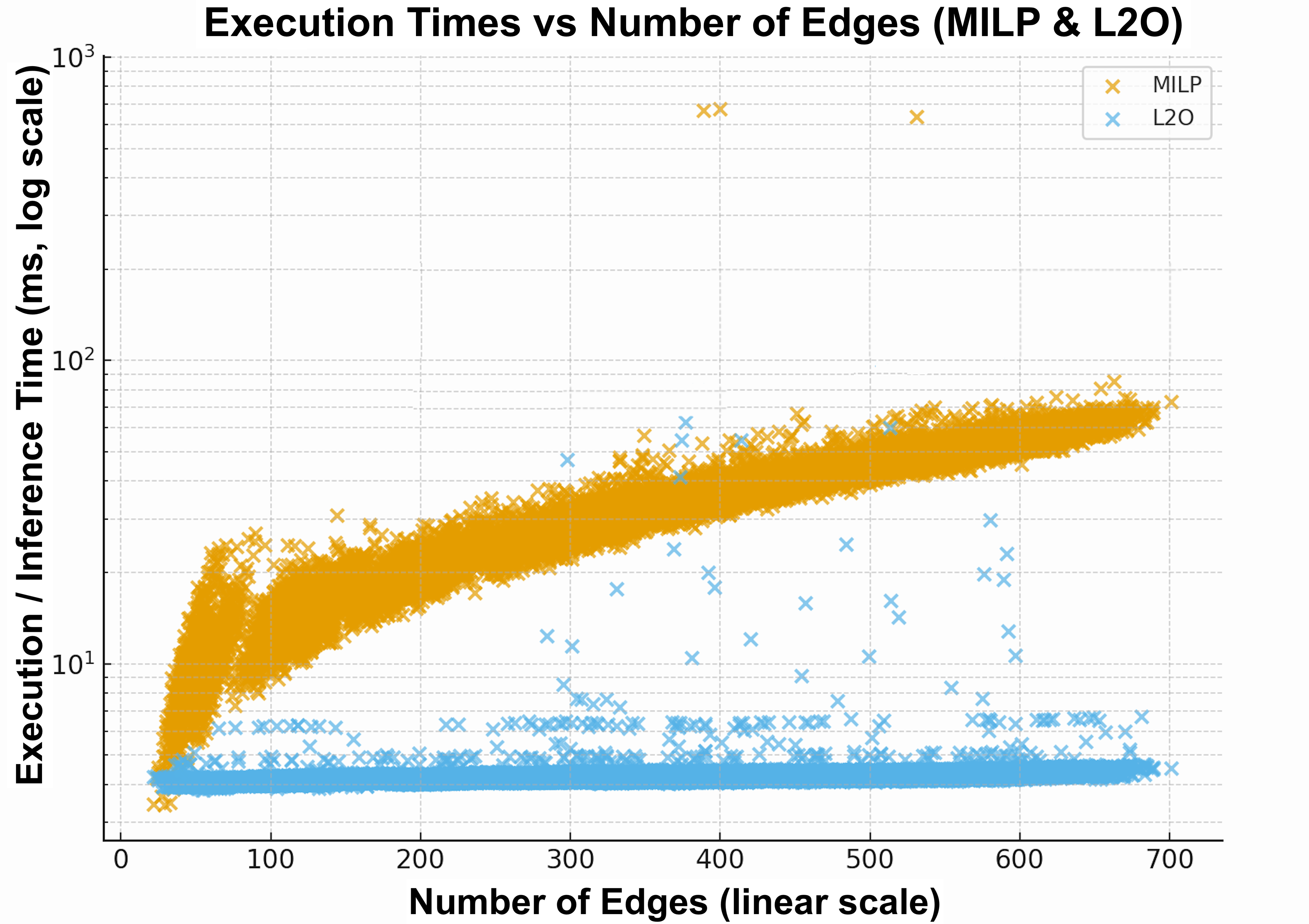}}
\caption{Execution time comparison between the MILP solver and the proposed GIN-based L2O framework.}
\label{figL2O}
\end{figure}

\subsection{Connectivity Gain}
To quantify the benefit of relay-assisted multi-hop communication in NR-V2X,
we evaluate the \emph{connectivity gain} achieved when CAVs are allowed to
reach the infrastructure through intermediate relay nodes. Connectivity is
defined as the ability of a CAV to establish a feasible path toward at least one
RSU, which serves as the uplink access point for infotainment, cooperative
perception, and high–data-rate services.

As a baseline, we consider a \emph{single-hop} V2I configuration in which each
CAV attempts to connect directly to the nearest RSU. Due to sparse RSU density,
urban obstructions, NLoS conditions, and variable link capacity, a non-negligible
fraction of vehicles remains disconnected.
Enabling \emph{multi-hop V2V relaying} allows additional CAVs to forward their
traffic through neighbouring vehicles, effectively extending the RSU coverage
footprint. The MILP oracle computes the optimal set of relay links that
maximises the number of connected CAVs under radio, capacity, and flow
constraints.

Across the entire dataset of 15{,}500 graph instances, the MILP solution yields
an average connectivity gain of
\begin{equation}
\Delta_{\text{conn}}^{\text{MILP}} = 9.31\%,
\end{equation}
representing the fraction of vehicles that become connected through optimal
multi-hop paths. Under the same conditions, the GIN-based L2O surrogate attains
\begin{equation}
\Delta_{\text{conn}}^{\text{L2O}} = 5.84\%,
\end{equation}
recovering a substantial portion of the optimal multi-hop benefit while
operating at real-time inference speed.
These improvements correspond to CAVs that would otherwise remain isolated in
a single-hop scenario, illustrating the tangible system-level advantage of
learning-based relay activation in realistic NR-V2X deployments.

\section{Conclusion}
This paper has introduced a GIN-based Learning-to-Optimise framework for
multi-hop relay-link activation in NR-V2X networks. The proposed method learns
to approximate optimal Mixed-Integer Linear Programming (MILP) solutions by
means of a supervised edge-classification strategy based on topology-aware
graph embeddings. By exploiting the expressiveness of Graph Isomorphism
Networks and the discriminative power of an edge-level MLP, the model is able
to capture both structural and radio-related properties of dynamic vehicular
topologies.
Extensive experiments over simulated urban V2X snapshots demonstrate
that the proposed approach achieves high edge-level accuracy relative to the
MILP oracle while significantly reducing computational latency. In addition,
L2O preserves the optimal connectivity gain obtained by multi-hop relaying in
dense urban scenarios, providing a practical alternative to repeated MILP
invocations.

Overall, the results confirm that GNN-driven Learning-to-Optimise techniques
constitute a promising direction for real-time resource management in the next-
generation intelligent transportation systems. Future work will extend the
proposed framework toward spatio-temporal models capable of sequential
decision-making under mobility, and investigate hybrid L2O–MILP policies for
real-time NR-V2X deployment.

%%%%%%%%%%

\bibliographystyle{IEEEtran}
\bibliography{References.bib}

\end{document}